\documentclass{IEEEtran} 
\IEEEoverridecommandlockouts

\usepackage{graphics}
\usepackage{epsfig}
\usepackage{times}
\usepackage{amsmath}
\usepackage{amssymb}
\usepackage{graphicx}
\usepackage{epstopdf}
\usepackage{epsfig}
\usepackage{enumerate}
\usepackage[noadjust]{cite}
\usepackage{color}
\usepackage{url}

\newtheorem{Theorem}{Theorem}
\newtheorem{Lemma}{Lemma}
\newtheorem{Definition}{Definition}
\newtheorem{Remark}{Remark}

\title{\LARGE \bf Privacy-Preserved Collaborative Estimation for Networked Vehicles with Application to Road Anomaly Detection}

\author{Huan~Gao, Zhaojian Li, Yongqiang Wang
	\thanks{The work was supported in part by the National Science Foundation under Grants 1824014 and 1912702.}
	\thanks{Huan Gao and Yongqiang Wang are with the Department of Electrical and Computer Engineering, Clemson University, Clemson, SC 29634, USA {\tt\small \{hgao2, yongqiw\}@clemson.edu}}
	\thanks{Zhaojian Li is with the Department of Mechanical Engineering, Michigan State University, East Lansing, MI 48824, USA {\tt\small lizhaoj1@egr.msu.edu}}
}

\begin{document}

\maketitle

\begin{abstract}
Road information such as road profile has been widely used in intelligent vehicle systems to improve road safety, ride comfort, and fuel economy. However, practical challenges, such as vehicle heterogeneity, parameter uncertainty, and measurement reliability, make it extremely difficult for a single vehicle to accurately and reliably estimate such information. To overcome these limitations, we propose a new learning-based collaborative estimation approach by fusing information from a fleet of networked vehicles. However, information exchange among these vehicles necessary for collaborative estimation may disclose sensitive information such as individual vehicle's identity, which poses serious privacy threats. To address this issue, we propose a unified privacy-preserving collaborative estimation framework which allows connected vehicles to iteratively refine estimation results through exploiting sequential measurements made by multiple vehicles traversing the same road segment. The collaborative estimation approach systematically incorporates privacy-protection schemes into the estimation design and exploits estimation dynamics to obscure exchanged information. Different from patching conventional privacy mechanisms like differential privacy that will compromise algorithmic accuracy or homomorphic encryption that will incur heavy communication/computation overhead, the dynamics enabled privacy protection does not sacrifice accuracy or significantly increase communication/computation overhead. Numerical simulations confirm the effectiveness of our proposed approach.
\end{abstract}

\section{Introduction}

With increasingly enhanced sensing capabilities on modern vehicles, there is a growing interest in employing road information such as black ice and road profile in intelligent vehicle systems to enhance road safety \cite{li2015road}, ride comfort \cite{li2014cloud}, and fuel economy \cite{ozatay2014cloud, hu2017cyber}. Real-time and crowd-sourced road information can increase situational awareness, enhance control performance, and provide new functionalities. For instance, road roughness and anomaly information has been used in comfort-based route planning \cite{li2016road} and suspension control \cite{li2014cloud}; road grade information has been shown to be able to improve fuel economy when integrated in powertrain control \cite{santin2017adaptive, krupadanam2013road}; and real-time road surface friction information can be used to enhance the performance of braking and steering control \cite{singh2015estimation}. Modern vehicles are equipped with a rich set of sensors that are readily available to be exploited to discover the aforementioned road information.

Vehicle interactions with road and traffic can be modeled as dynamical systems where road or traffic conditions such as grade, friction coefficient, road profile, and traffic density can be modeled as disturbances or system states. Therefore, input and state observers have been extensively used to estimate road and traffic information in automotive and transportation engineering in the past decades \cite{kidambi2014methods, singh2015estimation, li2016optimal, zhang2017hierarchical, doumiati2011estimation, lendek2010fuzzy, work2008ensemble}. For example, the authors in \cite{kidambi2014methods} modeled road grade as a system state and constructed a state observer to estimate it. Some other papers such as \cite{li2016optimal} modeled roadway velocity disturbances as system inputs and employed an input observer to estimate them.

However, almost all existing road information estimation approaches are developed in a single-vehicle setting, which renders the estimation result susceptible to vehicle variability, parameter uncertainty, and measurement reliability. To overcome such limitations, we propose to exploit multiple (heterogeneous) vehicles to cooperatively estimate road information with model-induced learning signals, relayed from earlier participating vehicles to subsequent vehicles, for enhanced performance. Furthermore, while the proposed collaborative estimation is expected to offer enhanced performance, the information exchange necessary for the implementation of the collaborative estimation may result in the disclosure of sensitive vehicle information and lead to privacy breaches. 

In fact, privacy of networked vehicles is not a new problem \cite{li2005caravan, hubaux2004security}. In recent years, the fact that V2V communications can transmit a vehicle's position raises serious concerns about position and identity privacy \cite{parkinson2017cyber} --- an adversary can use V2V communications to track a car without being noticed \cite{li2005caravan, dotzer2005privacy, corser2016evaluating}, leading to many potential malicious activities \cite{hubaux2004security}. To address the urgent need for privacy, plenty of V2V privacy-preserving approaches have been proposed based on conventional information technology privacy mechanisms such as cryptography \cite{yao1982protocols, shamir1979share, prabhakaran2013secure}, $k$-anonymity \cite{sweeney2002k}, differential privacy \cite{dwork2006calibrating}, or information-theoretical privacy \cite{sankar2013utility}. Recently results also emerged on the privacy preservation in vehicles based crowd-sourcing \cite{chim2012vspn, ni2016privacy}. However, these approaches need a mighty trusted central authority having access to the identity of all participants and are inappropriate for the scenario considered in this work for two reasons. First, such a central authority may not exist, particularly in large-scale systems like swarm robots. Secondly, even a central server exists, it may not be fully trustable, i.e., it may be honest-but-curious which follows all communications/computations correctly but is curious about users' private information.

In this paper, we propose a unified framework of privacy-preserving collaborative estimation to fuse local road estimation from a fleet of networked vehicles without leaking individual vehicles' sensitive information. More specifically, by leveraging vehicle-to-vehicle and vehicle-to-infrastructure communications, we develop a decentralized collaborative estimation framework for multiple vehicles traveling on the same road segment to iteratively refine the estimation results. In particular, building upon our prior work on single-vehicle based optimal state estimation \cite{li2016optimal}, we develop an iterative learning based estimation approach in which multiple vehicles sequentially relay their successive measurements of the same road information (e.g., road profile) to iteratively enhance collaborative estimation performance. Our framework emulates iterative learning control (ILC) that is frequently used to treat repetitive disturbances \cite{xu2008real} and tune the feed-forward control signal iteratively based on the memory data from previous iterations. It is worth noting that conventional ILC assumes that the system plant and its operations remain the same over iterations. However, for road information estimation, vehicles are inherently heterogeneous and hence existing ILC theory cannot be applied. Therefore, our collaborative estimation approach nontrivially generalizes the conventional ILC theory to allow collaborative estimation among a sequence of heterogeneous vehicles. To enable privacy preservation between participating vehicles, we directly incorporate privacy protection design in the collaborative estimation framework. More specifically, by leveraging the inherent dynamical properties of collaborative estimation, we enable the obfuscation of exchanged messages in a completely decentralized manner without sacrificing algorithmic accuracy or incurring heavy communication/computation overhead. Therefore, our privacy mechanism does not need a trusted central authority, which is different from most of existing approaches. Furthermore, compared with patching conventional privacy mechanisms like differential privacy which will compromise algorithmic accuracy or homomorphic encryption (as done in our prior work \cite{ruan2019secure, zhang2019admm}) which will incur heavy communication/computation overhead, our approach is superior in that it enables privacy protection in collaborative estimation without sacrificing accuracy or significantly increasing communication/computation overhead.

The remainder of this paper is organized as follows. Section II introduces the problem description and preliminary background on road information estimation in a single vehicle setting. A collaborative estimation framework for networked vehicles is proposed in Section III, which is followed by the dynamics-enabled privacy-preserving design in Section IV. Simulation experiments are presented in Section V. Finally we conclude this paper in Section VI.

\section{Problem Description and Preliminaries}

In this section, we present the problem description and review vehicle dynamics and our prior work on single vehicle-based estimation, which provides necessary context and the foundation for the proposed privacy-preserving collaborative estimation framework.

\subsection{Problem Description}
\begin{figure}
	\centering
	\vspace{0.5cm}\includegraphics[width=\linewidth]{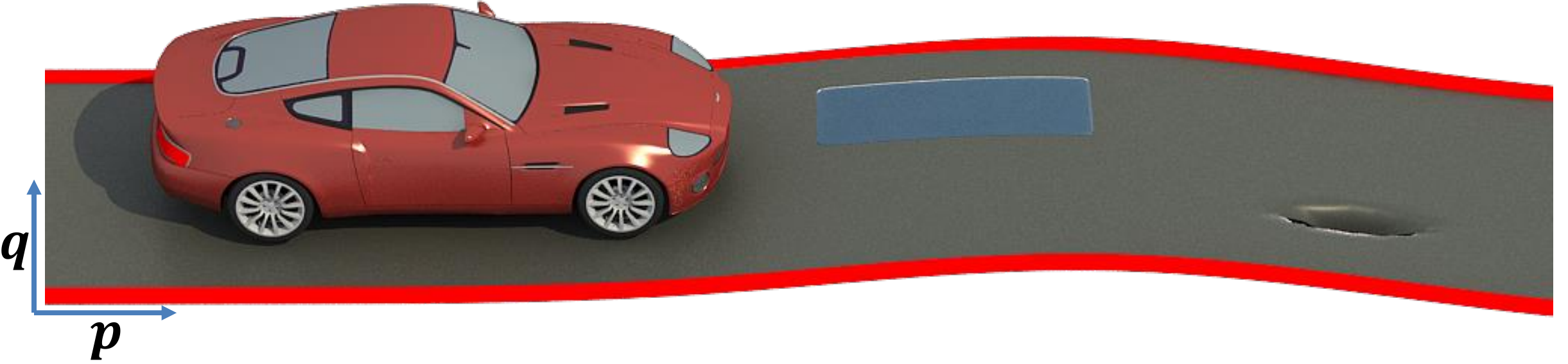}
	\caption{Illustration of a road segment with one pothole.}
	\label{fig_road}
\end{figure}

Road profile has been frequently proposed to be incorporated as a preview to enhance suspension controls for improved safety and comfort \cite{Preview1, Preview2, Preview3}. Given a road segment (e.g., defined by two consecutive road mile markers \cite{roadmarker}) as illustrated in Fig. \ref{fig_road}, the objective of vehicle-based road profile estimation is to use existing onboard sensors (e.g., accelerometers, yaw rate, roll rate, GPS) to discover road profile information, which can be characterized by $w(p)$, a function of distance in the longitudinal direction (the $p$ direction in Fig. \ref{fig_road}). By scaling the distance $p$ with the vehicle speed, the road information to be estimated can also be represented by $w(t)$, a function of time. Model-based road profile estimation approaches exploit onboard measurements along with the underlying dynamics to reconstruct $w$ as well as to estimate vehicle states for feedback control \cite{grade, Profile1}. We next introduce the underlying vehicle dynamics and an estimation framework to discover road profile in a single-vehicle setting.

\subsection{Vehicle Dynamical Model}

\begin{figure}[h]
	\begin{center}
		\vspace{0.2cm}\includegraphics[width=0.42\textwidth]{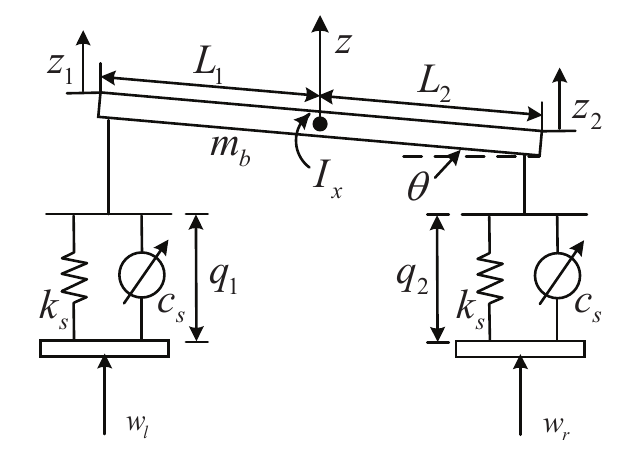}
	\end{center}
	\caption{A reduced front half-car model.}
	\label{fig_half_car}
\end{figure}

Model-based road information discovery relies on a model that characterizes the underlying dynamics of vehicle-road interaction. In this work, we consider a reduced front half-car model as shown in Fig. \ref{fig_half_car}. The front half car body is modeled as a rigid body with mass $m_b$. $I_x$ represents the moment of inertia about the longitudinal axis. The vertical displacement of the center of gravity (CG), left body tip, and right body tip, from equilibrium, are denoted by $z$, $z_1$, and $z_2$, respectively. $L_1$ and $L_2$ represent the left and right tip-to-CG distances, respectively. The parameters $k_s$ and $c_s$ represent the spring stiffness and damping coefficient of the suspension system, respectively. We further assume that the left and right sides have the same suspension parameters $k_s$ and $c_s$. We denote the roll angle by $\theta$. The variables $q_1$ and $q_2$ represent the left and right suspension deflections from equilibrium values, respectively. The signals $w_l$ and $w_r$ are the road velocity inputs to the left and right wheels, respectively. Since the wheels have high stiffness, we assume that $w_l$ and $w_r$ are directly applied to the left and right suspensions, respectively.

By defining $x_1=q_1$, $x_2=q_2$, $x_3=\dot{z}$, and $x_4=\dot{\theta}$ as the states, we have the following equations of motion
\begin{equation}\label{eqn_motion}
\begin{aligned}
\dot{x}_1 &= x_3+L_1x_4-w_l \\ 
\dot{x}_2 &= x_3-L_2x_4-w_r \\ 
m_b\dot{x}_3 &= -k_sx_1 - c_s\dot{x}_1 - k_sx_2 - c_s\dot{x}_2 \\ 
\frac{1}{2}I_x\dot{x}_4 &= -L_1k_sx_1 - L_1c_s\dot{x}_1 + L_2k_sx_2 + L_2c_s\dot{x}_2 \\ 
\end{aligned}
\end{equation}
Two sensor measurements, vertical accelerations of the left and right body tips, are used as outputs, which have the following form
\begin{equation}\label{eqn_half_car_measurements}
\begin{aligned}
y_1 & =\ddot{z}_1 =\ddot{z} + L_1 \ddot{\theta} = \dot{x}_3 + L_1 \dot{x}_4 \\
y_2 & =\ddot{z}_2 =\ddot{z} - L_2 \ddot{\theta} = \dot{x}_3 - L_2 \dot{x}_4
\end{aligned}
\end{equation}
Let ${x}=[x_1, \, x_2, \, x_3, \, x_4]^T$, ${y}=[\ddot{z}_1, \, \ddot{z}_2]^T$, ${w}=[w_l, \, w_r]^T$, and ${v}=[v_1, \, v_2]^T$ represent the state vector, measurement vector, input vector, and measurement noise vector, respectively, we can rewrite (\ref{eqn_motion}) and (\ref{eqn_half_car_measurements}) in a compact matrix form as follows
\begin{equation}\label{eqn_motion_state}
\begin{aligned}
\dot{{x}} &= {A} {x} + {B} {w}\\ 
{y} &= {C} {x} +{D} {w} + {v} \\ 
\end{aligned}
\end{equation}
where ${A}$, ${B}$, ${C}$, and ${D}$ are constant matrices derived from (\ref{eqn_motion}) and (\ref{eqn_half_car_measurements}).

\subsection{Road Information Estimation Based on a Single Vehicle}

In this subsection, we introduce some preliminary results on road information estimation in a single vehicle setting, which includes an input observer \cite{kalabic2013multi} and a state estimator driven by JDP (jump-diffusion process) \cite{li2016optimal}.

\subsubsection{Input observer}
To estimate the road input ${w}$, we employ an input observer proposed by \cite{kalabic2013multi}, which is given as follows
\begin{equation}\label{eqn_input_observer}
\begin{aligned}
\dot{{\varepsilon}} &= - \gamma {S} {\varepsilon} + \gamma {S} {K} {A} {x} + (\gamma {S})^2 {K} {x}\\
\hat{{w}} &= -{\varepsilon} + \gamma {S} {K} {x} 
\end{aligned}
\end{equation}
where ${\varepsilon}$ is the observer state, $\gamma > 0.5$ is a scalar gain, and ${S}=0.5(1+\gamma){I}_2$, and ${K}=({B}^T{B})^{-1}{B}^T$.

\subsubsection{Jump-diffusion process based optimal state estimator}

Note that to employ the input observer in (\ref{eqn_input_observer}), the state information ${x}$ is required. Therefore, a state estimator is needed to estimate ${x}$ from measurements ${y}$ in (\ref{eqn_motion_state}). Historically, diffusion (or Wiener) processes have been used to model stochastic disturbances in the development of state estimation methodologies, which have been applied in road information discovery such as road grade estimation \cite{kidambi2014methods} and tire-road friction estimation \cite{singh2015estimation}. However, rare but pronounced events that can induce significant impact (such as a car hitting potholes or speed bumps) may be better modeled as jumps (Poisson processes). Therefore, jump-diffusion process (JDP), involving both jumps and diffusions, can be used to model road disturbances to a car \cite{kolmanovsky2006stochastic, li2016optimal}. More specifically, we denote the road input $w$ as ${w}= \dot{{\eta}} + {\sigma_{\zeta}} \dot{{\zeta}}$, where ${\eta}$ is a vector jump process with each component having the same Poisson parameter $\lambda$ and ${\zeta}$ is a standard vector Wiener process with $\sigma_{\zeta} \sigma_{\zeta}^T$ being its covariance. The processes $\eta$ and $\zeta$ are assumed to be independent. Furthermore, the jump size of each component of $\eta$ is also a random variable. We denote the jump size mean and covariance by $\mu_{\eta}$ and $\Sigma_{\eta}$, respectively. A JDP-based state estimator was developed in our prior work in \cite{li2016optimal} as follows:
\begin{equation}\label{eqn_state_estimator}
\begin{aligned}
\dot{\hat{{x}}} &= {A} \hat{{x}} + {F}({C}\hat{{x}} - {y}) + ({B} + {F}{D}) \lambda {\mu}_{{\eta}}
\end{aligned}
\end{equation}
where ${F}$ is the estimator gain to be determined.
\begin{Lemma}\label{lemma_state_estimator}
	(cf. Theorem 1 of \cite{li2016optimal}). Suppose the pair $({A}, \, {C})$ is detectable, the pair $({A}, \, {B})$ is stabilizable, and ${S}^T {S} > 0$. Then, the optimal gain ${F}$ that minimizes
	\begin{equation}\label{eqn_estimator_error}
	\begin{aligned}
	J= \lim\limits_{t \rightarrow \infty} \frac{1}{t} E \int_{0}^{t} \big({x}(\tau)-\hat{{x}}(\tau)\big)^T {S}^T {S} \big({x}(\tau)-\hat{{x}}(\tau)\big) d\tau
	\end{aligned}
	\end{equation}
	in the open set of all gains ${F}$ for which $({A} + {F} {C})$ is asymptotically stable (Hurwitz) is given by
	\begin{equation}\label{eqn_estimator_gain}
	\begin{aligned}
	{F}= -{B}{\bar{\Sigma}}{D}^T {V}_2^{-1} - {Q}{C}^T{V}_2^{-1}
	\end{aligned}
	\end{equation}
	where ${Q}$ is the unique positive semi-definite solution to
	\begin{equation}\label{eqn_solution_Q}
	\begin{aligned}
	&({A} - {B}{\bar{\Sigma}}{D}^T {V}_2^{-1}{C}) {Q} + {Q} ({A} - {B}{\bar{\Sigma}}{D}^T {V}_2^{-1}{C})^T \quad \ \ \\
	& \qquad \qquad \qquad \qquad \qquad \quad + {V}_1 - {Q}{C}^T{V}_2^{-1}{C}{Q} = {0}
	\end{aligned}
	\end{equation}
	In (\ref{eqn_solution_Q}), ${\bar{\Sigma}} ={\sigma_{\zeta}} {\sigma_{\zeta}}^T + \lambda {\mu}_{{\eta}} {\mu}_{{\eta}}^T + \lambda {\Sigma}_{{\eta}}$, ${V}_1 = {B}{\bar{\Sigma}}{B}^T - {B}{\bar{\Sigma}}{D}^T {V}_2^{-1}{D}{\bar{\Sigma}}{B}^T$, and ${V}_2={\sigma_{\zeta}} {\sigma_{\zeta}}^T + {D}{\bar{\Sigma}}{D}^T$.
\end{Lemma}

\section{Collaborative Estimation for Networked Vehicles}

In this section, we extend the single vehicle-based estimator in Section II and develop a collaborative estimation framework by employing sequential measurements from multiple heterogeneous vehicles. Specifically, our collaborative estimation approach is inspired by iterative learning control (ILC). In particular, using sequential measurements taken from multiple vehicles traveling on the same road segment, we develop a completely decentralized iterative collaborative estimation approach by leveraging communication between vehicles which can be achieved using vehicle-to-vehicle or vehicle-to-infrastructure communications. 

The proposed collaborative estimation framework is illustrated in Fig. \ref{fig_collaborative_estimation}. Consider a road segment (e.g., defined by road mileposts). Let $j$ represent the sequence number of vehicles that drive over the road segment and participate in the collaborative estimation. The iterative learning framework exploits sequential estimation error $e_j$ ($j=1,2,\cdots$) and learning signal $w_j^f$ ($j=1,2,\cdots$) to iteratively refine the road information estimate. More specifically, let $P_j$ represent the plant of vehicle $j$ and $\hat P_j$ be the plant model. Also let $D_j$ represent the local input observer, $y_j$ be the measurements, and $\hat{y}_j=\hat P\hat w_j$ be the output by feeding the estimated input $\hat w_j$ to the plant model. Further let $T_j$ denote the dynamics from the true road disturbance $w_j$ to the estimation error $e_j$ and $S_j$ denote the dynamics from the learning signal $w_j^f$ to $e_j$. From Fig. \ref{fig_collaborative_estimation}, we have $T_j=P_j-\hat P_jD_jP_j$ and $S_j=-\hat P_j$. The proposed collaborative estimation framework is summarized in Algorithm 1 below:
\begin{figure*}[h]
	\begin{center}
		\includegraphics[width=1\textwidth]{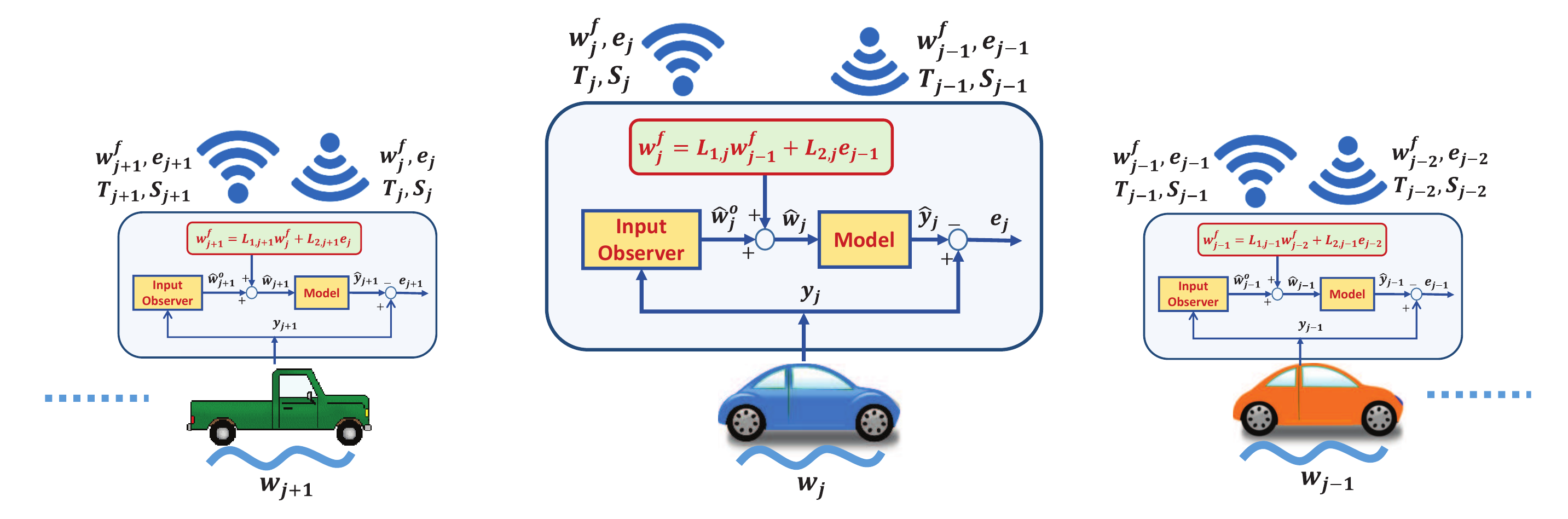}
	\end{center}
	\caption{Schematic diagram of iterative learning based collaborative estimation framework. A learning signal $w_j^f$ is injected to vehicle $j$ to relay estimation errors from heterogeneous vehicles.}
	\label{fig_collaborative_estimation}
\end{figure*}

\noindent\rule{0.49\textwidth}{0.5pt}
\noindent\textbf{Algorithm 1: Proposed collaborative estimation framework}

\vspace{-0.2cm}\noindent\rule{0.49\textwidth}{0.5pt}
A fleet of networked vehicles travel on the same road segment and participate in the collaborative estimation of the road input $w$. For each vehicle $j$:
\begin{enumerate}
	\item After traversing the considered road segment, vehicle $j$ collects its measurements $y_{j}$. 
	\item Using the measurements $y_{j}$, vehicle $j$ employs the JDP-based state estimator in (\ref{eqn_state_estimator}) to estimate its state $x$. Based on the estimate of the state $x$, vehicle $j$ uses the input observer in (\ref{eqn_input_observer}) to get an initial estimate of the road input $w$, which is denoted as $\hat{w}_{j}^o$.
	\item After receiving information $T_{j-1}$, $S_{j-1}$, $e_{j-1}$, and $w_{j-1}^f$ from vehicle $j-1$, vehicle $j$ constructs its learning filters $L_{1,j}$ and $L_{2,j}$ (the way how to construct learning filters will be discussed below), and then obtains its learning signal $w_j^f = L_{1,j}w_{j-1}^f + L_{2,j}e_{j-1}$.
	\item Using $\hat{w}_{j}= \hat{w}_{j}^o + w_{j}^f$, vehicle $j$ obtains its estimate of the road input.
	\item By feeding the estimated signal $\hat{w}_{j}$ to the plant model $\hat{P}_{j}$, vehicle $j$ obtains the signal $\hat{y}_{j}$ and the error signal $e_{j}$ using $e_{j}=y_{j}-\hat{y}_{j}$.
	\item Vehicle $j$ sends its information $T_{j}$, $S_{j}$, $e_{j}$, and $w_{j}^f$ to vehicle $j+1$.
\end{enumerate}
\vspace{-0.2cm}\rule{0.49\textwidth}{0.5pt}

From the proposed collaborative estimation framework we know that vehicle $j$ receives the error signal $e_{j-1}$ and learning signal $w_{j-1}^f$ from an earlier participating vehicle $j-1$. So the inputs to the estimation system on vehicle $j$ are the true road disturbance $w_j$ and the learning signal $w_j^f$. The measurement-prediction mismatch $e_j$ can be neatly described as
\begin{equation}\label{eqn_e_j}
e_j=T_jw_j+S_jw_j^f
\end{equation}
where $T_j$ and $S_j$ are transfer functions from $w_j$ to $e_j$ and from $w_j^f$ to $e_j$, respectively, as defined above. Note that they can be represented in either time domain in the form of state space or frequency domain using transfer functions. We design the following iterative estimation mechanism for vehicle $j$
\begin{equation}\label{eqn_w_j}
w_j^f = L_{1,j}w_{j-1}^f + L_{2,j}e_{j-1}
\end{equation}
where $L_{1,j}$ and $L_{2,j}$ are the learning filters to be designed, $w_{j-1}^f$ is the learning signal relayed from vehicle $j-1$, and $e_{j-1}$ is the error signal from vehicle $j-1$. Plugging (\ref{eqn_w_j}) into (\ref{eqn_e_j}) leads to
\begin{equation}\label{eqn_e_j_decomposition}
\begin{aligned}
e_j = & T_jw_j+S_j\left[L_{1,j}w_{j-1}^f + L_{2,j}e_{j-1}\right] \\
= & T_jw_j + S_jL_{2,j}e_{j-1} + S_jL_{1,j}S_{j-1}^{-1} \left[e_{j-1} - T_{j-1}w_{j-1}\right] \\
= & \left[S_jL_{2,j} + S_jL_{1,j}S_{j-1}^{-1}\right] e_{j-1} \\
& \quad + \left[T_j - S_jL_{1,j}S_{j-1}^{-1}T_{j-1}\right] w
\end{aligned}
\end{equation}
where we assumed $w_{j}=w_{j-1}=w,\forall j=1,2,\cdots$.

It is worth noting that (\ref{eqn_e_j_decomposition}) represents the dynamic estimation error from vehicle $j-1$ to vehicle $j$. Next we will design learning filters $L_{1,j}$ and $L_{2,j}$, such that ${||S_jL_{2,j}+S_jL_{1,j}S_{j-1}^{-1}||}$ and ${||T_j-S_jL_{1,j}S_{j-1}^{-1}T_{j-1}||}$ are minimized, i.e.,
\begin{equation}\label{eqn_L1_L2}
\min_{L_{1,j},\, L_{2,j} } ||S_jL_{2,j} + S_jL_{1,j}S_{j-1}^{-1} || + ||T_j - S_jL_{1,j}S_{j-1}^{-1}T_{j-1}||
\end{equation}

Noting that $||T_j-S_jL_{1,j}S_{j-1}^{-1}T_{j-1}||$ only depends on $L_{1,j}$, we can solve (\ref{eqn_L1_L2}) in a sequential manner. By setting 
\begin{equation}\label{eqn_L1}
L^*_{1,j}=S_j^{-1}T_jT_{j-1}^{-1}S_{j-1}
\end{equation}
we have $||T_j-S_jL^*_{1,j}S_{j-1}^{-1}T_{j-1}||=0$. Given $L^*_{1,j}=S_j^{-1}T_jT_{j-1}^{-1}S_{j-1}$, setting
\begin{equation}\label{eqn_L2}
L^*_{2,j}=-S_j^{-1}T_jT_{j-1}^{-1}
\end{equation}
leads to $||S_jL^*_{2,j}+S_jL^*_{1,j}S_{j-1}^{-1}||=0$.

It is worth noting that the proposed iterative collaborative estimation approach removes the fundamental assumption of identical vehicles in traditional ILC via explicitly taking vehicle dynamics into account. Therefore, the proposed approach is a highly nontrivial generalization to ILC since it overcomes the homogeneous assumption in ILC, which is extremely significant in practice because homogeneous vehicles rarely exist in practical transportation systems. With increasing availability of communication in networked vehicles, the proposed collaborative estimation framework can greatly enhance estimation performance. 

Note that under the proposed collaborative estimation framework, vehicle $j$ needs to send the error signal $e_{j}$, learning signal $w_{j}^f$, and its dynamics $T_{j}$ and $S_{j}$ to vehicle $j+1$ after driving through the road segment, so that vehicle $j+1$ can construct learning filters $L_{1,j+1}$ and $L_{2,j+1}$, and further get its learning signal $w_{j+1}^f$ using (\ref{eqn_w_j}). However, such information exchange necessary for the implementation of the collaborative estimation may result in the disclosure of sensitive vehicle information and lead to privacy breaches. In the next section, we will present a new privacy-preserving design to address such privacy concerns.

\section{Privacy Protection in the Collaborative Estimation Framework}

As analyzed in Section III, the proposed collaborative estimation framework requires explicitly exchanging of $T_{j}$ and $S_{j}$, which contain sensitive model and dynamics information of vehicle $j$. Such information could be used by a malicious party to infer sensitive information of vehicle $j$ such as its type and even identity and hence poses serious privacy threats. In the considered collaborative estimation environment, the malicious party could be a vehicle participating in the collaborative estimation or it could be an external eavesdropper wiretapping communication channels. Therefore, before putting the sensitive information $T_{j}$ and $S_{j}$ out on the communication channel and sending them to vehicle $j+1$, vehicle $j$ has to obfuscate and cover such information. It is worth noting that anonymity has been discussed to enable privacy of V2V communications by making the message sender indistinguishable. However, anonymity has its limitations such as increased computational latency/communication overhead and reduced scalability \cite{hubaux2004security}. Moreover, as indicated in \cite{le2014differentially}, simple anonymization is usually not enough to guarantee privacy as privacy breaches generally arise from the possibility of linking anonymized data with public side information, as demonstrated in \cite{narayanan2008robust,calandrino2011you}. Furthermore, in existing anonymization based privacy-preserving approaches, it is assumed that a trusted server exists to manage the real identifies of all participants in order to track and isolate malicious participants \cite{hubaux2004security, wu2009balanced}. Such an assumption may not be valid in practice. It is also worth noting that other conventional privacy-preserving approaches like differential privacy or homomorphic encryption (as adopted in our prior work \cite{ruan2019secure, zhang2019admm}) are not desirable here since differential privacy unavoidably compromises the accuracy of computation and homomorphic encryption incurs heavy communication/computation overhead.

Given that the exchanged information $T_{j}$ and $S_{j}$ contain the sensitive model and dynamics information of vehicle $j$, we propose a new privacy design seamlessly integrated with our collaborative estimation framework. To this end, we first give our attack models and a precise definition of privacy.
\begin{itemize}
	\item \emph{Eavesdropping attacks} are attacks in which an external eavesdropper wiretaps communication channels to intercept exchanged messages in an attempt to learn the information about sending vehicles.
	\item \emph{Honest-but-curious attacks} are attacks in which attackers follow all protocol steps correctly but are curious and collect all received intermediate data in an attempt to learn the information about other participating vehicles.
\end{itemize}

\begin{Definition}
	The privacy of vehicle $j$ is preserved if an attacker cannot infer the dynamics $T_{j}$ and $S_{j}$ of vehicle $j$. More specifically, attackers cannot infer the exact zeros and poles of $T_{j}$ and $S_{j}$.
\end{Definition}

We propose to exploit the inherent dynamical properties of collaborative estimation to obfuscate exchanged information. More specifically, instead of sending $T_{j}$, $S_{j}$, $e_{j}$, and $w_{j}^f$ directly from vehicle $j$ to vehicle $j+1$, we propose to let vehicle $j$ send 
\begin{equation}\label{eqn_obfuscate_info}
\begin{aligned}
\tilde{T}_{j} & =\Psi^{T1}_{j} T_{j} \Psi^{T2}_{j}\\
\tilde{S}_{j} & =\Psi^{S1}_{j} S_{j} \Psi^{S2}_{j}\\
\tilde{e}_{j} & =\Psi^e_{j} e_{j} \\ 
\tilde{w}^f_{j} & =\Psi^w_{j} w^f_{j}
\end{aligned}
\end{equation}
instead, where $\Psi^{T1}_{j}$, $\Psi^{T2}_{j}$ $\Psi^{S1}_{j}$, $\Psi^{S2}_{j}$, $\Psi^e_{j}$ and $\Psi^w_{j}$ are obfuscating dynamical systems, i.e., $\Psi^{T1}_{j}(s)$, $\Psi^{T2}_{j}(s)$, $\Psi^{S1}_{j}(s)$, $\Psi^{S2}_{j}(s)$, $\Psi^e_{j}(s)$, and $\Psi^w_{j}(s)$ are generated by and only known to vehicle $j$. Note that since $e_{j}$ and $w_{j}^f$ are two-dimensional signals and $T_{j}$ and $S_{j}$ are MIMO transfer functions, the obfuscating dynamical systems should also be matrices of appropriate dimensions.

The difficulties in designing obfuscating dynamical systems lie in eliminating their influence on the accuracy of collaborative estimation, i.e., in guaranteeing the optimality of $L_{1,j+1}$ and $L_{2,j+1}$ for the update of $w_{j+1}^f$. We prove that if the obfuscating dynamics are designed according to Theorem \ref{theorem_privacy_accuracy}, then the optimality of $L_{1,j+1}$ and $L_{2,j+1}$ will not be affected at all, i.e., the collaborative estimation accuracy will not be affected at all by the privacy design:
\begin{Theorem}\label{theorem_privacy_accuracy}
	The information obfuscation framework has no influence on the accuracy of collaborative estimation if the obfuscating dynamical systems satisfy the following relationships:
	\begin{equation}\label{eqn_obfuscate_relation}
	\begin{aligned}
	\Psi^{T1}_{j} & = \Psi^e_{j} = \Psi^{S1}_{j}\\
	\Psi^{T2}_{j} & = I \\
	\Psi^w_{j} & = (\Psi^{S2}_{j})^{-1}
	\end{aligned}
	\end{equation}
\end{Theorem}

\textit{Proof}: To prove that the information obfuscation framework has no influence on the accuracy of collaborative estimation, it is sufficient to prove that the information obfuscation framework does not affect vehicle $j+1$'s computing accuracy of $w_{j+1}^f$ since the information from vehicle $j$ only involves in the computation of $w_{j+1}^f$.

We first show how the obfuscating dynamical systems in (\ref{eqn_obfuscate_relation}) affect the design of $L_{1,j+1}$ and $L_{2,j+1}$ on vehicle $j+1$. Under the information obfuscation framework, vehicle $j+1$ designs $L_{1,j+1}$ in (\ref{eqn_L1}) and $L_{2,j+1}$ in (\ref{eqn_L2}) as follows
\begin{equation}\label{eqn_tilde_L1}
\begin{aligned}
\tilde{L}^*_{1,j+1} & = S_{j+1}^{-1} T_{j+1} \tilde{T}_{j}^{-1} \tilde{S}_{j} \\
& = S_{j+1}^{-1} T_{j+1} (\Psi^{T1}_{j} T_{j} \Psi^{T2}_{j})^{-1} \Psi^{S1}_{j} S_{j} \Psi^{S2}_{j} \\
& = S_{j+1}^{-1} T_{j+1} T^{-1}_{j} S_{j} \Psi^{S2}_{j} \\
& = L^{*}_{1,j+1} \Psi^{S2}_{j}
\end{aligned}
\end{equation}
and
\begin{equation}\label{eqn_tilde_L2}
\begin{aligned}
\tilde{L}^*_{2,j+1} & =-S_{j+1}^{-1} T_{j+1} \tilde{T}_{j}^{-1} \\
& = -S_{j+1}^{-1} T_{j+1} (\Psi^{T1}_{j} T_{j} \Psi^{T2}_{j})^{-1} \\
& = -S_{j+1}^{-1} T_{j+1} T^{-1}_{j} (\Psi^{T1}_{j})^{-1} \\
& = L^*_{2,j+1} (\Psi^{T1}_{j})^{-1}
\end{aligned}
\end{equation}
Note that in the above derivation, we used $\Psi^{T1}_{j} =\Psi^{S1}_{j}$ and $\Psi^{T2}_{j} = I$ in (\ref{eqn_obfuscate_relation}). It is also worth noting that $\tilde{L}^*_{1,j+1}$ and $\tilde{L}^*_{2,j+1}$ are the optimal solution to (\ref{eqn_L1_L2}) under the information obfuscation framework since they lead to
\begin{equation}
||T_{j+1} - S_{j+1} \tilde{L}^*_{1,j+1} \tilde{S}_{j}^{-1} \tilde{T}_{j}|| = 0
\end{equation}
and
\begin{equation}
||S_{j+1} \tilde{L}^*_{2,j+1} + S_{j+1} \tilde{L}^*_{1,j+1} \tilde{S}_{j}^{-1}||=0
\end{equation}

Next we evaluate the influence of the information obfuscation on $w_{j+1}^f$:
\begin{equation}\label{eqn_tilde_w_f}
\tilde{w}_{j+1}^f=\tilde{L}^*_{1,j+1}\tilde{w}_{j}^f+\tilde{L}^*_{2,j+1}\tilde{e}_{j}
\end{equation}
Plugging (\ref{eqn_tilde_L1}) and (\ref{eqn_tilde_L2}) into
(\ref{eqn_tilde_w_f}) leads to
\begin{equation}
\begin{aligned}
\tilde{w}_{j+1}^f & = L^{*}_{1,j+1} \Psi^{S2}_{j} \Psi^w_{j} w_{j}^f + L^*_{2,j+1} (\Psi^{T1}_{j})^{-1} \Psi^e_{j} e_{j} \\
& = L^{*}_{1,j+1} w_{j}^f + L^*_{2,j+1} e_{j} \\
& = w_{j+1}^f
\end{aligned}
\end{equation}
where we used $\Psi^w_{j}=(\Psi^{S2}_{j})^{-1}$ and $\Psi^{T1}_{j}=\Psi^e_{j}$ in (\ref{eqn_obfuscate_relation}). 

Therefore, we can see that once the obfuscating dynamical systems satisfy the relationships in (\ref{eqn_obfuscate_relation}), the information obfuscation mechanism will not affect the computation of $w_{j+1}^f$ and further $\hat{w}_{j+1}$ on vehicle $j+1$, meaning that the information obfuscation framework has no influence on the accuracy of collaborative estimation. \hfill{$\blacksquare$}

Combining (\ref{eqn_obfuscate_info}) and (\ref{eqn_obfuscate_relation}) leads to
\begin{equation}\label{eqn_obfuscate_message}
\begin{aligned}
\tilde{T}_{j} & =\Psi^{S1}_{j} T_{j}\\
\tilde{S}_{j} & =\Psi^{S1}_{j} S_{j} \Psi^{S2}_{j}\\
\tilde{e}_{j} & =\Psi^{S1}_{j} e_{j} \\ 
\tilde{w}^f_{j} & =(\Psi^{S2}_{j})^{-1} w^f_{j}
\end{aligned}
\end{equation}
So vehicle $j$ only needs to design $\Psi^{S1}_{j}$ and $\Psi^{S2}_{j}$ to obfuscate its private dynamics $T_{j}$ and $S_{j}$. Incorporating (\ref{eqn_obfuscate_message}) in the collaborative estimation framework, we propose the information obfuscation mechanism as follows:

\noindent\rule{0.49\textwidth}{0.5pt}
\noindent\textbf{Algorithm 2: Privacy-protection mechanism}

\vspace{-0.2cm}\noindent\rule{0.49\textwidth}{0.5pt}
In Algorithm 1, vehicle $j$ replaces Step 6) with the following steps:
\begin{enumerate}
	\item Vehicle $j$ randomly chooses two positive integers $n_1$ and $n_2$, and keeps them private to itself.
	\item Vehicle $j$ randomly selects $n_1$ poles (from the left-half of the $s$-plane) and four groups of zeros with each group having $n_1$ zeros. Using one group of zeros and $n_1$ poles, vehicle $j$ can construct a transfer function. Therefore, using these four groups of zeros and $n_1$ poles, vehicle $j$ constructs a 2-by-2 transfer function matrix as $\Psi^{S1}_{j}$. $\Psi^{S1}_{j}$ will be private to vehicle $j$.
	\item Vehicle $j$ randomly selects $n_2$ poles (from the left-half of the $s$-plane) and four groups of zeros with each group having $n_2$ zeros. Using these four groups of zeros and $n_2$ poles, vehicle $j$ constructs a 2-by-2 transfer function matrix as $\Psi^{S2}_{j}$. $\Psi^{S2}_{j}$ will be private to vehicle $j$.
	\item Using the obfuscating dynamical systems $\Psi^{S1}_{j}$ and $\Psi^{S2}_{j}$, vehicle $j$ obfuscates its $T_{j}$, $S_{j}$, $e_{j}$, and $w^f_{j}$ according to (\ref{eqn_obfuscate_message}), and then sends the obfuscated information $\tilde{T}_{j}$, $\tilde{S}_{j}$, $\tilde{e}_{j}$, and $\tilde{w}^f_{j}$ to vehicle $j+1$.
\end{enumerate}
\vspace{-0.2cm}\rule{0.49\textwidth}{0.5pt}

Next we show that our proposed information obfuscation mechanism can indeed achieve the defined privacy.

\begin{Theorem}\label{theorem_privacy_performance}
	Our privacy-preserving mechanism can protect the privacy of vehicle $j$'s dynamics $T_{j}$ and $S_{j}$ using the obfuscating dynamical systems $\Psi^{S1}_{j}$ and $\Psi^{S2}_{j}$.
\end{Theorem}

\textit{Proof}: To prove that the privacy of vehicle $j$ can be protected, it is sufficient to prove that $T_{j}$ and $S_{j}$ cannot be inferred by an attacker. Our idea is to prove that neither honest-but-curious vehicle $j+1$ nor an eavesdrop attacker can distinguish whether the original dynamics of vehicle $j$ is $T_{j}$ (resp. $S_{j}$) or $\bar{T}_{j}$ (resp. $\bar{S}_{j}$) where $\bar{T}_{j}$ and $\bar{S}_{j}$ can be any stable dynamics that have different orders, zeros, and poles from $T_{j}$ and $S_{j}$, respectively. Under the information obfuscation framework, we assume that any attacker has access to the obfuscated dynamics $\tilde{T}_{j}$ and $\tilde{S}_{j}$ sent by vehicle $j$. Therefore, if we can prove that under any stable dynamics $\bar{T}_{j}$ and $\bar{S}_{j}$, the obfuscated dynamics $\tilde{T}_{j}$ and $\tilde{S}_{j}$ could keep unchanged, then neither honest-but-curious vehicle $j+1$ nor an eavesdrop attacker can infer the original dynamics $T_{j}$ and $S_{j}$ (including their zeros and poles).

It can be proven that under any stable dynamics $\bar{T}_{j}$ and $\bar{S}_{j}$, there always exist obfuscating dynamical systems 
\begin{equation}
\begin{aligned}
\bar{\Psi}^{S1}_{j} & = \Psi^{S1}_{j} T_{j} \bar{T}_{j}^{-1} \\
\bar{\Psi}^{S2}_{j} & = \bar{S}_{j}^{-1} \bar{T}_{j} T_{j}^{-1} S_{j} \Psi^{S2}_{j} \\
\end{aligned}
\end{equation}
making the obfuscated dynamics $\tilde{T}_{j}$ and $\tilde{S}_{j}$ exactly the same as under the original dynamics $T_{j}$ and $S_{j}$, i.e.,
\begin{equation}
\begin{aligned}
\tilde{T}_{j} & =\Psi^{S1}_{j} T_{j} = \bar{\Psi}^{S1}_{j} \bar{T}_{j}\\
\tilde{S}_{j} & =\Psi^{S1}_{j} S_{j} \Psi^{S2}_{j} = \bar{\Psi}^{S1}_{j} \bar{S}_{j} \bar{\Psi}^{S2}_{j}\\
\end{aligned}
\end{equation}
Therefore, neither honest-but-curious vehicle $j+1$ nor an eavesdrop attacker can infer the original dynamics $T_{j}$ and $S_{j}$, meaning that our privacy-preserving approach can protect the privacy of vehicle $j$'s dynamics $T_{j}$ and $S_{j}$ using the obfuscating dynamical systems $\Psi^{S1}_{j}$ and $\Psi^{S2}_{j}$. \hfill{$\blacksquare$}


\begin{Remark}
	The design of $\Psi^{S1}_{j}$ and $\Psi^{S2}_{j}$ is subject to a trade-off between complexity and performance. That is, to provide a stronger privacy protection, it is desirable to use higher-order transfer functions $\Psi^{S1}_{j}$ and $\Psi^{S2}_{j}$ having complicated dynamics (more zeros and poles) to cover the original models. However, a higher order of the dynamics of $\Psi^{S1}_{j}$ and $\Psi^{S2}_{j}$ will also unavoidably improve the computational complexity. In fact, as can be seen from our proof in Theorem \ref{theorem_privacy_performance}, as long as $\Psi^{S1}_{j}$ and $\Psi^{S2}_{j}$ have one pole and zero (private to itself), an attacker will be unable to infer the true dynamics of vehicle $j$.
\end{Remark}

\begin{Remark}
	The signals $e_{j}$ and $w_{j}^f$ will also be completely reshaped and covered by the obfuscating dynamical systems $\Psi^{S1}_{j}$ and $\Psi^{S2}_{j}$ since different frequency components of $e_{j}$ and $w_{j}^f$ will be amplified/attenuated differently by $\Psi^{S1}_{j}$ and $\Psi^{S2}_{j}$. Therefore, information of the original signals $e_{j}$ and $w_{j}^f$ can also be covered by the information obfuscation framework.
\end{Remark}

\begin{Remark}
	Different from existing privacy-preserving approaches which rely on encryption or pseudonyms management, the proposed privacy-preserving approach employs self-generated obfuscating dynamical systems which are simple in computation, lightweight in communication, and completely scalable in implementation. Furthermore, compared with differential privacy based approaches which add additive noise to exchanged signals and hence unavoidably affect algorithmic accuracy, our approach does not sacrifice the accuracy of estimation, which is crucial in safety-critical dynamical systems.
\end{Remark}

\section{Simulation Results}

\begin{figure}[h]
	\begin{center}
		\includegraphics[width=0.5\textwidth]{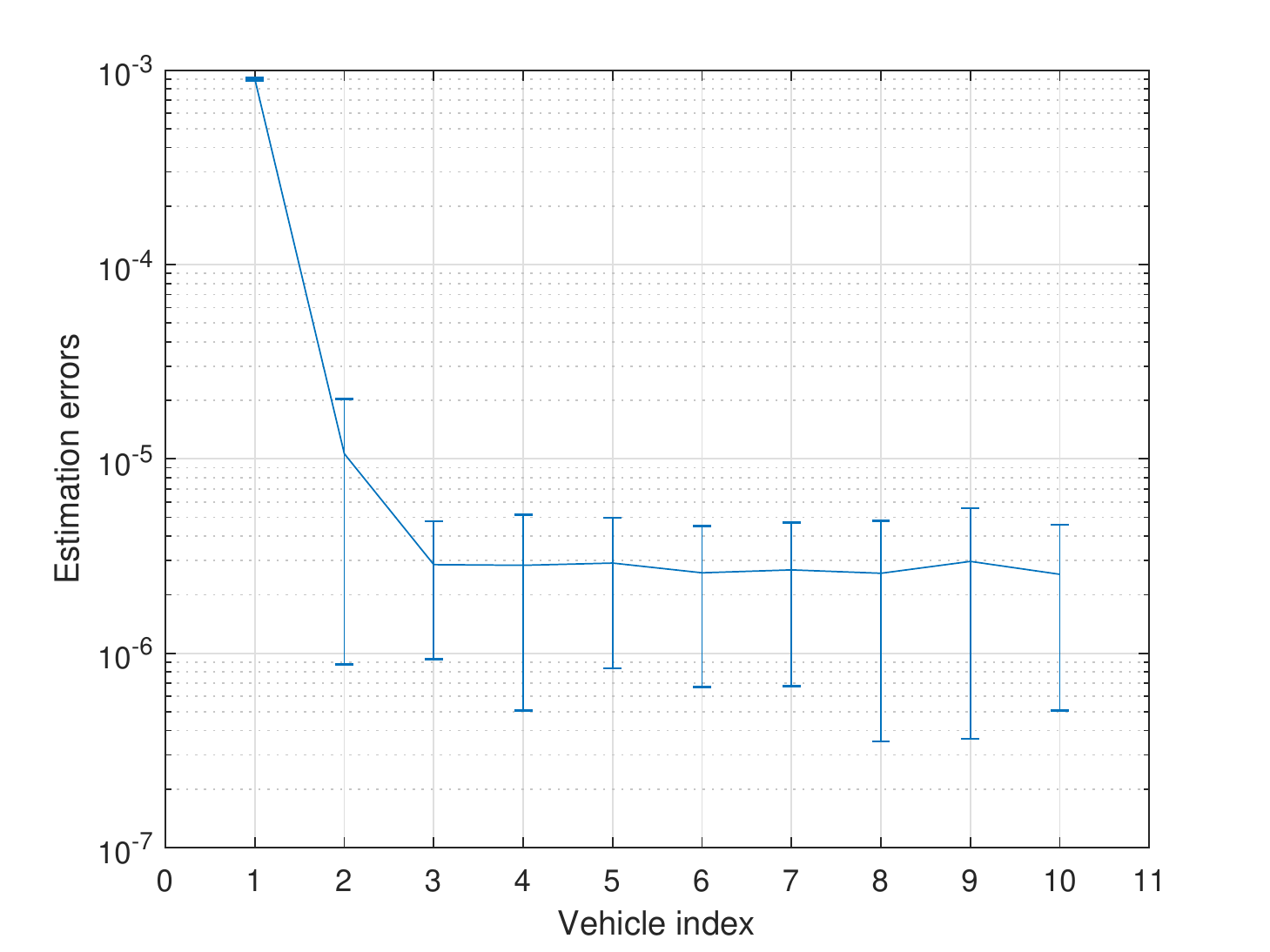}
	\end{center}
	\caption{Estimation performance of a sequence of $10$ heterogeneous vehicles.}
	\label{fig_estimation_performance}
\end{figure}

Following our proposed collaborative estimation approach, a sequence of $10$ heterogeneous vehicles were simulated to collaboratively improve the estimation performance. We assume that the exact values of these vehicle parameters such as $m_b$, $k_s$, $c_s$, and $I_x$ are not available, and we can only access unbiased estimations of these parameters. We repeat the simulation for $100$ times, and record the estimation errors in terms of mean square error for each simulation. The averaged estimation performance is shown in Fig. \ref{fig_estimation_performance}, from which we can see that the estimation performance is significantly better than single vehicle based estimation (compare vehicle $3$ and vehicle $1$ --- note that the error bar in Fig. \ref{fig_estimation_performance} represents the standard deviation of estimation errors for each vehicles in the $100$ simulations and if only one vehicle is used in traditional estimation, then its performance will be the same as vehicle $1$ in our framework.)

\begin{figure}[!h]
	\begin{center}
		\includegraphics[width=0.5\textwidth]{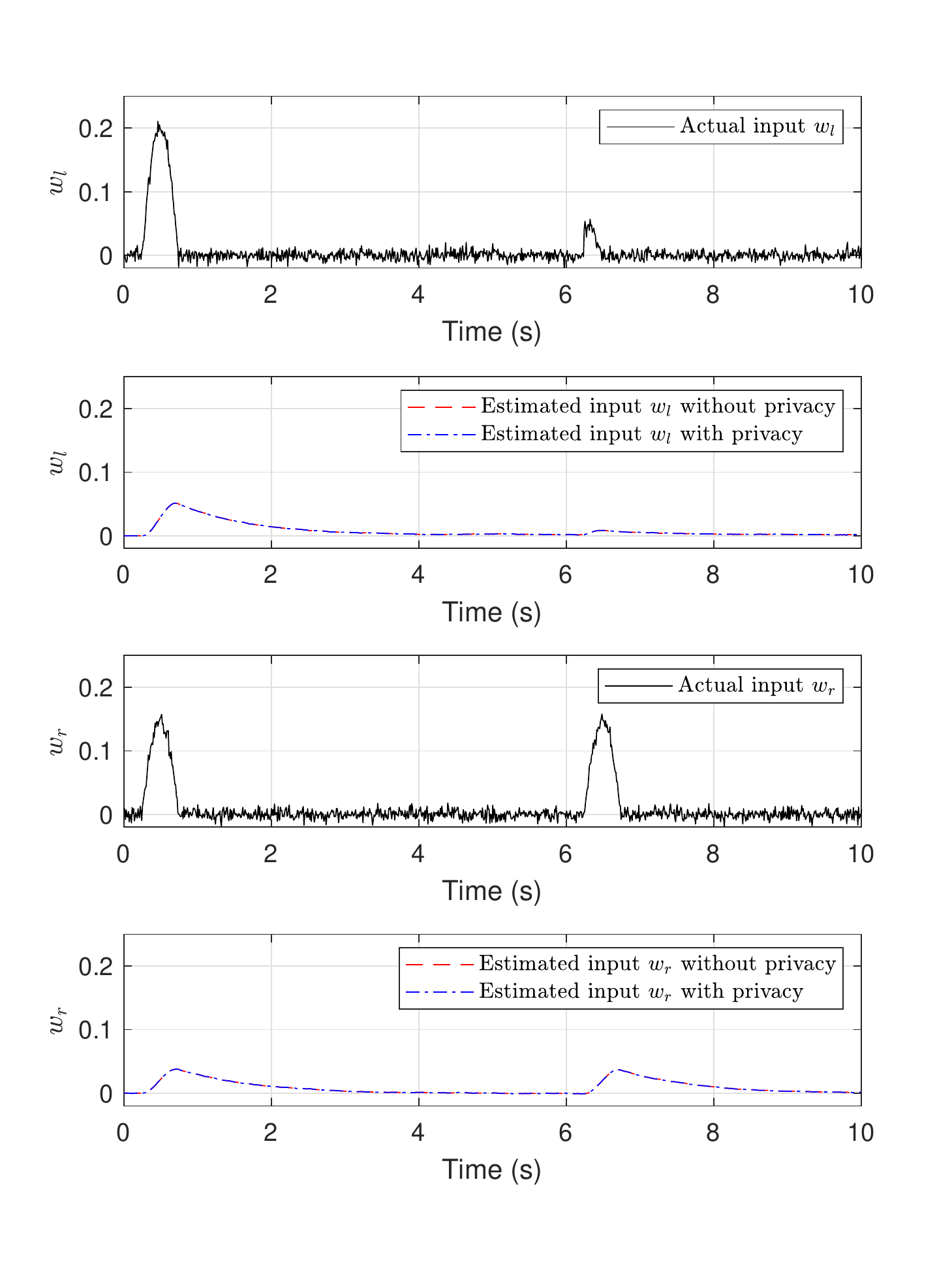}
	\end{center}
	\vspace{-0.8cm}\caption{Comparisons of the actual road input signals with the estimated signals using and not using our privacy-preserving design for vehicle $1$.}
	\label{fig_comparison_vehicle_1}
\end{figure}

\begin{figure}[h]
	\begin{center}
		\includegraphics[width=0.5\textwidth]{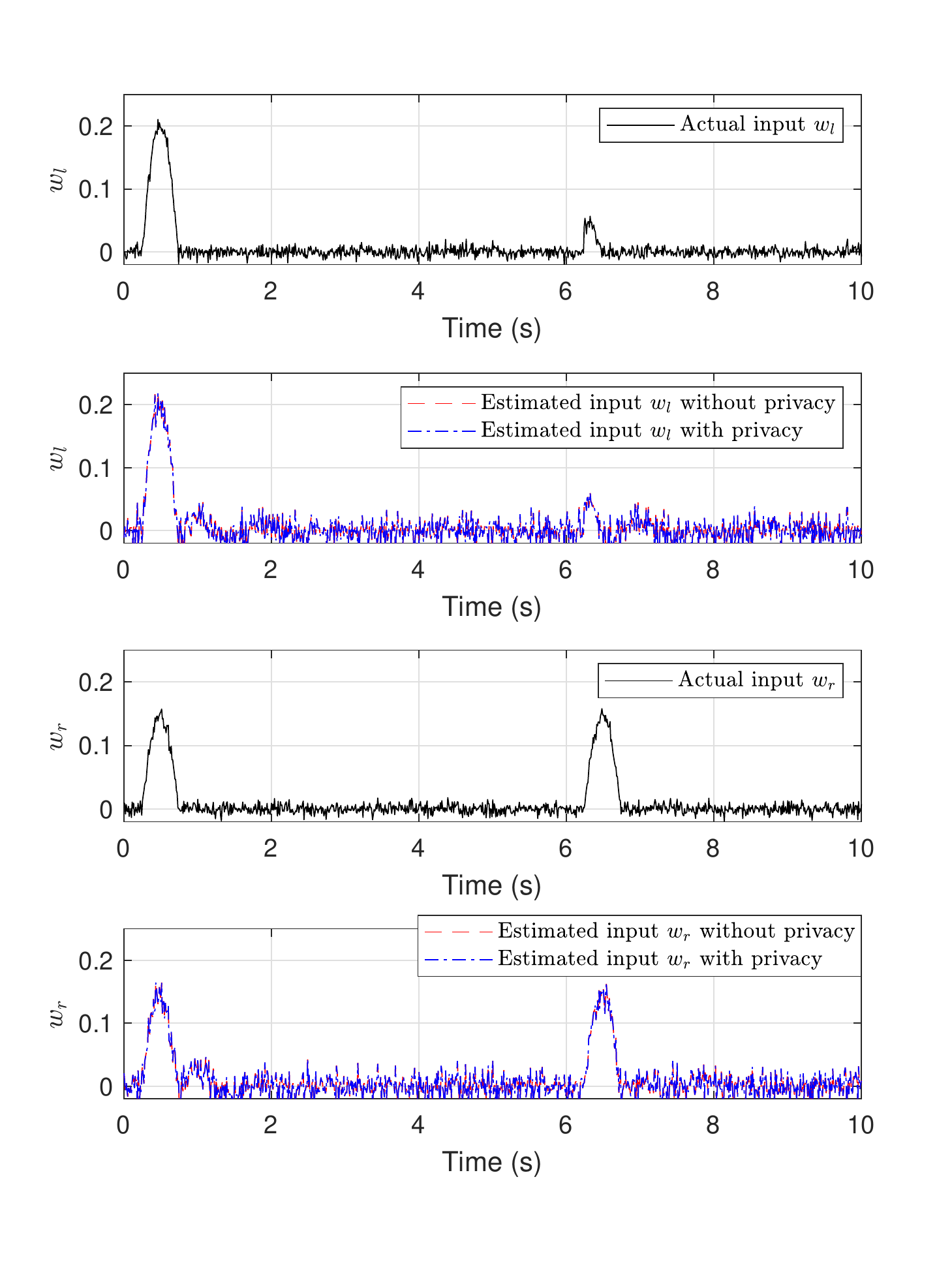}
	\end{center}
	\vspace{-0.8cm}\caption{Comparisons of the actual road input signals with the estimated signals using and not using our privacy-preserving design for vehicle $2$.}
	\label{fig_comparison_vehicle_2}
\end{figure}

\begin{figure}[h]
	\begin{center}
		\includegraphics[width=0.5\textwidth]{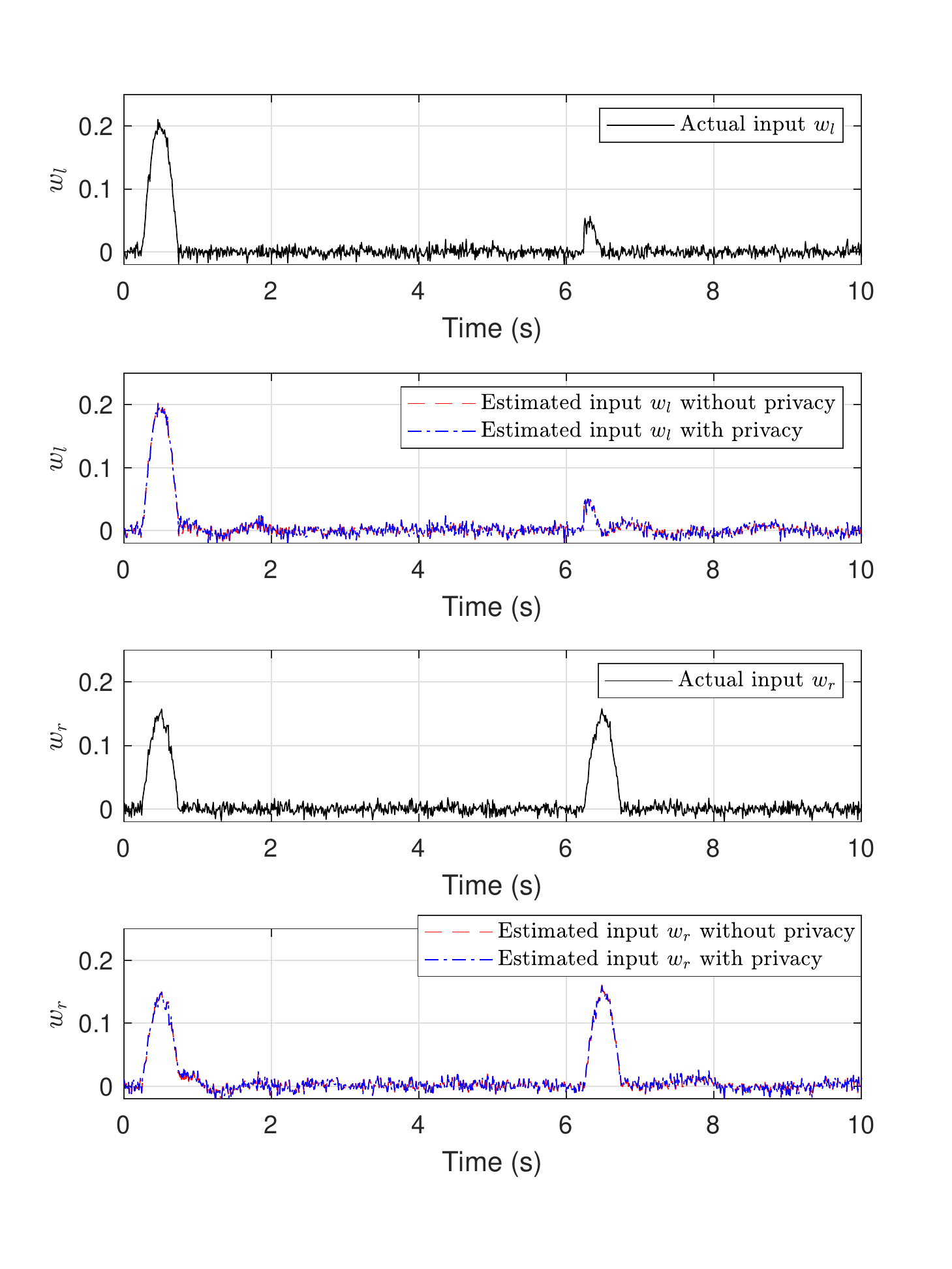}
	\end{center}
	\vspace{-0.8cm}\caption{Comparisons of the actual road input signals with the estimated signals using and not using our privacy-preserving design for vehicle $3$.}
	\label{fig_comparison_vehicle_3}
\end{figure}

We also perform simulations to show that our information obfuscation framework can protect the privacy of vehicle dynamics without compromising the accuracy of computation results. To this end, we first compare the actual road input signals with the estimated signals with and without implementing our privacy-preserving design. The comparison results for vehicles $1$, $2$, $3$ are presented in Fig. \ref{fig_comparison_vehicle_1}, Fig. \ref{fig_comparison_vehicle_2}, and Fig. \ref{fig_comparison_vehicle_3}, respectively. From the simulation results, we can see that our information obfuscation framework has no influence on the collaborative estimation accuracy since the estimated road inputs using our privacy-preserving design are the same as the ones without privacy design.


\begin{figure}[h]
	\begin{center}
		\includegraphics[width=0.5\textwidth]{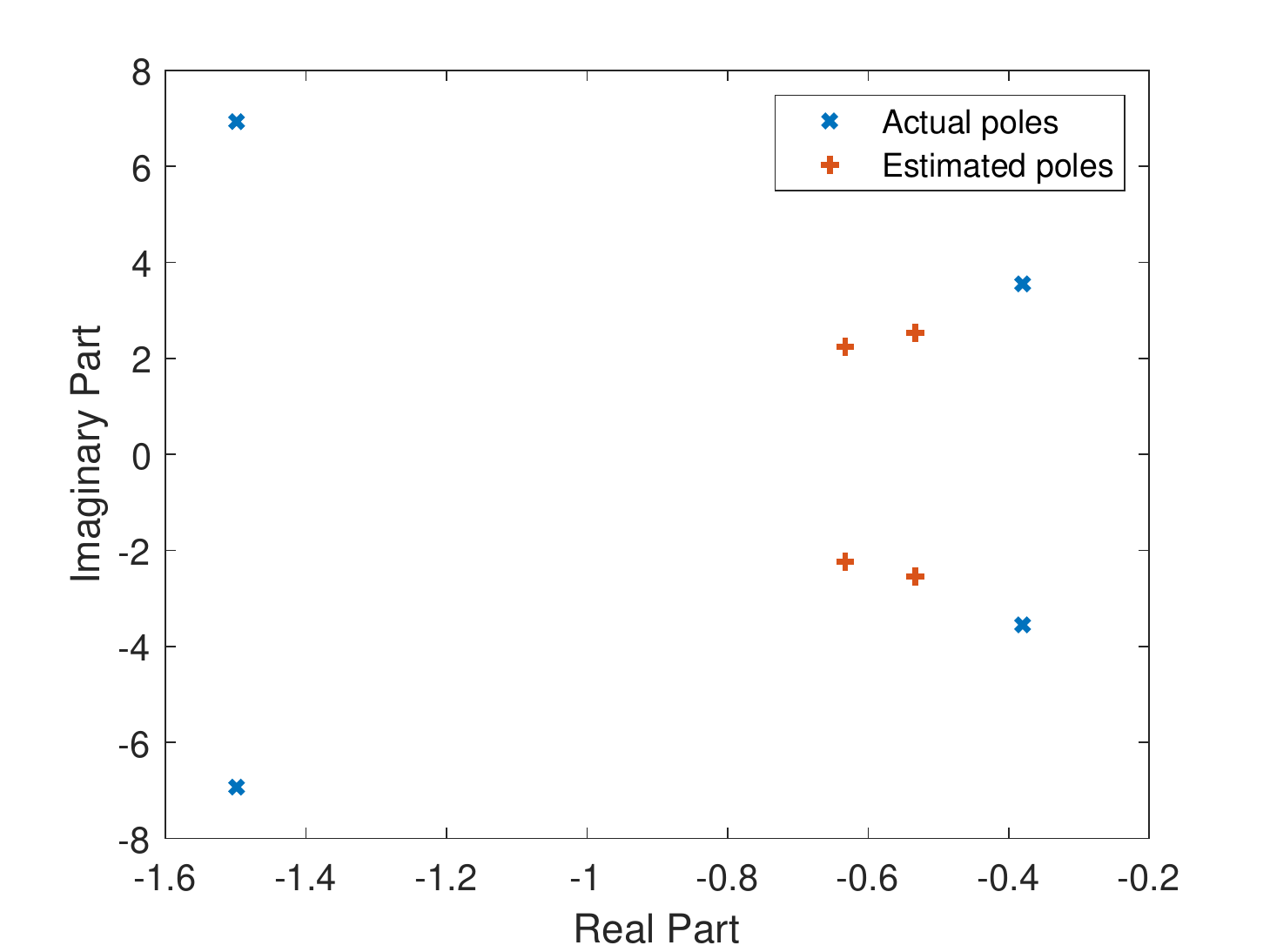}
	\end{center}
	\caption{Comparisons of the actual poles with the estimated poles of $S_{1}$ for vehicle $1$.}
	\label{fig_comparison_S_poles_1}
\end{figure}

\begin{figure}[h]
	\begin{center}
		\includegraphics[width=0.5\textwidth]{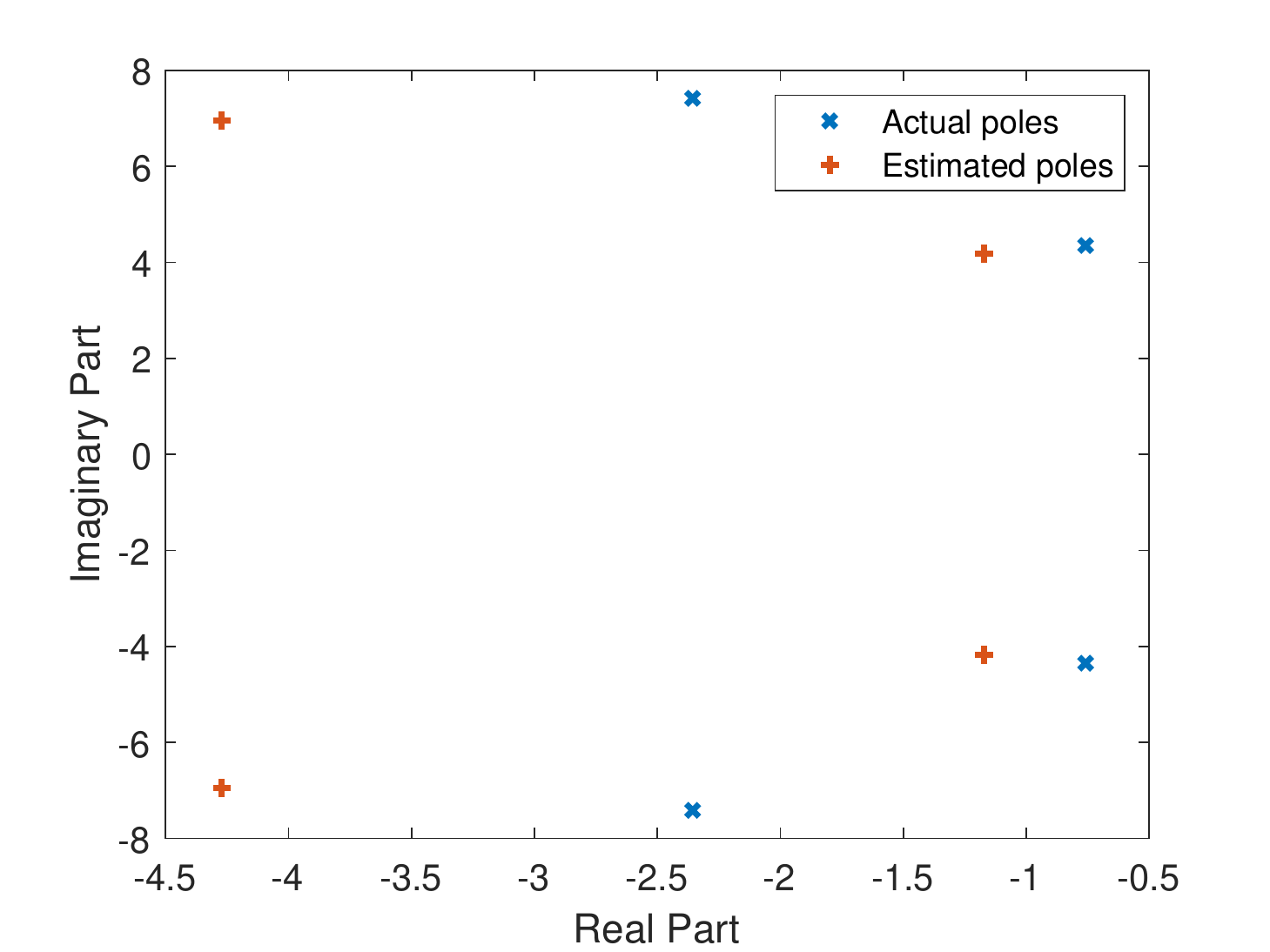}
	\end{center}
	\caption{Comparisons of the actual poles with the estimated poles of $S_{2}$ for vehicle $2$.}
	\label{fig_comparison_S_poles_2}
\end{figure}

\begin{figure}[h]
	\begin{center}
		\includegraphics[width=0.5\textwidth]{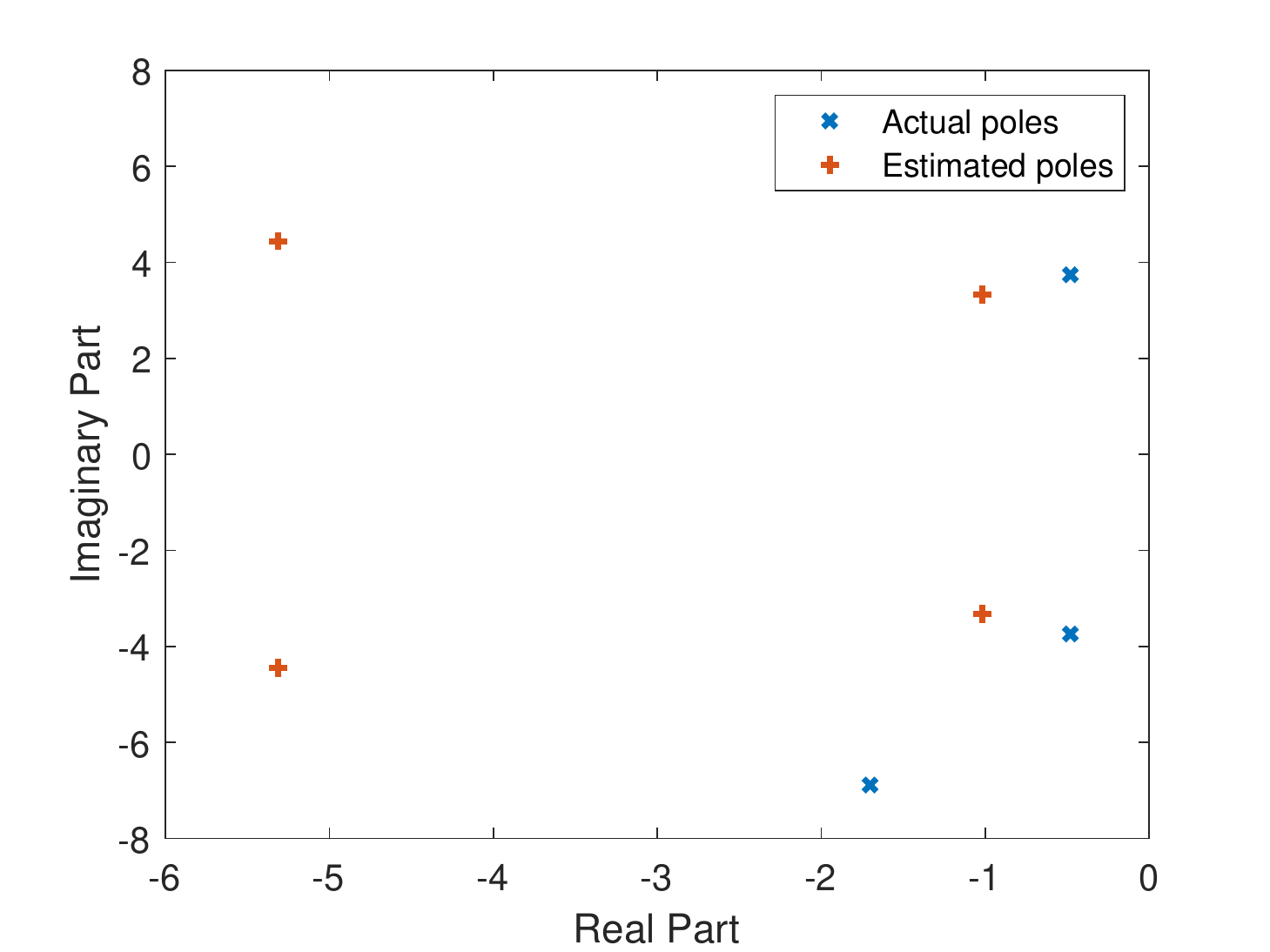}
	\end{center}
	\caption{Comparisons of the actual poles with the estimated poles of $S_{3}$ for vehicle $3$.}
	\label{fig_comparison_S_poles_3}
\end{figure}

We then evaluate the privacy performance of our information obfuscation mechanism. As analyzed in Section IV, instead of sending the sensitive dynamics $T_{j}$ and $S_{j}$, vehicle $j$ sends obfuscated dynamics $\tilde{T}_{j}$ and $\tilde{S}_{j}$ which have different orders, zeros, and poles from $T_{j}$ and $S_{j}$. We assume that an attacker knows the orders of dynamics $T_{j}$ and $S_{j}$, and intends to infer the actual zeros and poles of $T_{j}$ and $S_{j}$ based on received dynamics $\tilde{T}_{j}$ and $\tilde{S}_{j}$. As the attacker knows the actual orders, it reduces the orders of $\tilde{T}_{j}$ and $\tilde{S}_{j}$ to the orders of $T_{j}$ and $S_{j}$ and then infers the zeros and poles of vehicle $j$. In our simulation, the Model Reducer App in Matlab was used to reduce the order of a dynamical system. The comparison results between the actual poles and estimated poles of $S_{1}$, $S_{2}$, and $S_{3}$ are shown in Fig. \ref{fig_comparison_S_poles_1}, Fig. \ref{fig_comparison_S_poles_2}, and Fig. \ref{fig_comparison_S_poles_3}, respectively. From the simulation results we can see that the attacker cannot have a good estimate of the poles of dynamics $S_{j}$.

\section{Conclusions}

In this paper we proposed a unified framework for privacy-preserving collaborative estimation to fuse local road estimation from a fleet of networked vehicles. By generalizing the iterative learning control (ILC) technique, we established a novel collaborative estimation framework to enable heterogeneous vehicles to iteratively refine the estimation performance in a completely decentralized manner. Numerical simulations showed that the collaborative estimation approach can significantly enhance estimation performance compared with existing single-vehicle based estimation approaches. Given the importance of privacy protection in networked vehicles, we also developed a new privacy enabling mechanism which was seamlessly integrated in the collaborative estimation approach. By leveraging the inherent dynamical properties of collaborative estimation to obfuscate exchanged messages, our privacy-preserving design can be implemented in a completely decentralized manner without affecting the estimation accuracy or incurring heavy communication/computation overhead. Numerical simulations were provided to confirm the effectiveness of our proposed framework.

\bibliographystyle{unsrt}
\bibliography{abbr_bibli}

\begin{thebibliography}{10}

\bibitem{li2015road}
Zhaojian Li, Ilya Kolmanovsky, Ella Atkins, Jianbo Lu, Dimitar~P Filev, and
  John Michelini.
\newblock Road risk modeling and cloud-aided safety-based route planning.
\newblock {\em IEEE transactions on cybernetics}, 46(11):2473--2483, 2015.

\bibitem{li2014cloud}
Zhaojian Li, Ilya Kolmanovsky, Ella Atkins, Jianbo Lu, Dimitar Filev, and John
  Michelini.
\newblock Cloud aided semi-active suspension control.
\newblock In {\em 2014 IEEE Symposium on Computational Intelligence in Vehicles
  and Transportation Systems (CIVTS)}, pages 76--83. IEEE, 2014.

\bibitem{ozatay2014cloud}
Engin Ozatay, Simona Onori, James Wollaeger, Umit Ozguner, Giorgio Rizzoni,
  Dimitar Filev, John Michelini, and Stefano Di~Cairano.
\newblock Cloud-based velocity profile optimization for everyday driving: A
  dynamic-programming-based solution.
\newblock {\em IEEE Transactions on Intelligent Transportation Systems},
  15(6):2491--2505, 2014.

\bibitem{hu2017cyber}
Xiaosong Hu, Hong Wang, and Xiaolin Tang.
\newblock Cyber-physical control for energy-saving vehicle following with
  connectivity.
\newblock {\em IEEE Transactions on industrial electronics}, 64(11):8578--8587,
  2017.

\bibitem{li2016road}
Zhaojian Li, Ilya~V Kolmanovsky, Ella~M Atkins, Jianbo Lu, Dimitar~P Filev, and
  Yuchen Bai.
\newblock Road disturbance estimation and cloud-aided comfort-based route
  planning.
\newblock {\em IEEE transactions on cybernetics}, 47(11):3879--3891, 2016.

\bibitem{santin2017adaptive}
Ondrej Santin, Jaroslav Beran, Jaroslav Pekar, John Michelini, Junbo Jing,
  Steve Szwabowski, and Dimitar Filev.
\newblock Adaptive nonlinear model predictive cruise controller: Trailer tow
  use case.
\newblock Technical report, SAE Technical Paper, 2017.

\bibitem{krupadanam2013road}
Ashish~S Krupadanam, Mike~M McDonald, and William~C Albertson.
\newblock Road grade coordinated engine control systems, December~10 2013.
\newblock US Patent 8,606,483.

\bibitem{singh2015estimation}
Kanwar~Bharat Singh and Saied Taheri.
\newblock Estimation of tire--road friction coefficient and its application in
  chassis control systems.
\newblock {\em Systems Science \& Control Engineering}, 3(1):39--61, 2015.

\bibitem{kidambi2014methods}
Narayanan Kidambi, RL~Harne, Yuji Fujii, Gregory~M Pietron, and KW~Wang.
\newblock Methods in vehicle mass and road grade estimation.
\newblock {\em SAE International Journal of Passenger Cars-Mechanical Systems},
  7(2014-01-0111):981--991, 2014.

\bibitem{li2016optimal}
Zhaojian Li, Ilya~V Kolmanovsky, Uro{\v{s}}~V Kalabi{\'c}, Ella~M Atkins,
  Jianbo Lu, and Dimitar~P Filev.
\newblock Optimal state estimation for systems driven by jump--diffusion
  process with application to road anomaly detection.
\newblock {\em IEEE Transactions on Control Systems Technology},
  25(5):1634--1643, 2016.

\bibitem{zhang2017hierarchical}
Xudong Zhang and Dietmar G{\"o}hlich.
\newblock A hierarchical estimator development for estimation of tire-road
  friction coefficient.
\newblock {\em PLoS one}, 12(2):e0171085, 2017.

\bibitem{doumiati2011estimation}
Moustapha Doumiati, Alessandro Victorino, Ali Charara, and Daniel Lechner.
\newblock Estimation of road profile for vehicle dynamics motion: experimental
  validation.
\newblock In {\em Proceedings of the 2011 American control conference}, pages
  5237--5242. IEEE, 2011.

\bibitem{lendek2010fuzzy}
Zs~Lendek, R~Babu{\v{s}}ka, and B~De~Schutter.
\newblock Fuzzy models and observers for freeway traffic state tracking.
\newblock In {\em Proceedings of the 2010 American Control Conference}, pages
  2278--2283. IEEE, 2010.

\bibitem{work2008ensemble}
Daniel~B Work, Olli-Pekka Tossavainen, S{\'e}bastien Blandin, Alexandre~M
  Bayen, Tochukwu Iwuchukwu, and Kenneth Tracton.
\newblock An ensemble kalman filtering approach to highway traffic estimation
  using gps enabled mobile devices.
\newblock In {\em 2008 47th IEEE Conference on Decision and Control}, pages
  5062--5068. IEEE, 2008.

\bibitem{li2005caravan}
Mingyan Li, R~Poovendran, K~Sampigethaya, and L~Huang.
\newblock Caravan: Providing location privacy for vanet.
\newblock In {\em Proceedings of the Embedded Security in Cars (ESCAR)
  Workshop}, volume~2, pages 13--15, 2005.

\bibitem{hubaux2004security}
Jean-Pierre Hubaux, Srdjan Capkun, and Jun Luo.
\newblock The security and privacy of smart vehicles.
\newblock {\em IEEE Security \& Privacy}, 2(3):49--55, 2004.

\bibitem{parkinson2017cyber}
Simon Parkinson, Paul Ward, Kyle Wilson, and Jonathan Miller.
\newblock Cyber threats facing autonomous and connected vehicles: Future
  challenges.
\newblock {\em IEEE transactions on intelligent transportation systems},
  18(11):2898--2915, 2017.

\bibitem{dotzer2005privacy}
Florian D{\"o}tzer.
\newblock Privacy issues in vehicular ad hoc networks.
\newblock In {\em International Workshop on Privacy Enhancing Technologies},
  pages 197--209. Springer, 2005.

\bibitem{corser2016evaluating}
George~P Corser, Huirong Fu, and Abdelnasser Banihani.
\newblock Evaluating location privacy in vehicular communications and
  applications.
\newblock {\em IEEE transactions on intelligent transportation systems},
  17(9):2658--2667, 2016.

\bibitem{yao1982protocols}
Andrew~C Yao.
\newblock Protocols for secure computations.
\newblock In {\em 23rd annual symposium on foundations of computer science
  (sfcs 1982)}, pages 160--164. IEEE, 1982.

\bibitem{shamir1979share}
Adi Shamir.
\newblock How to share a secret.
\newblock {\em Communications of the ACM}, 22(11):612--613, 1979.

\bibitem{prabhakaran2013secure}
Manoj~M Prabhakaran and Amit Sahai.
\newblock {\em Secure multi-party computation}, volume~10.
\newblock IOS press, 2013.

\bibitem{sweeney2002k}
Latanya Sweeney.
\newblock $k$-anonymity: A model for protecting privacy.
\newblock {\em International Journal of Uncertainty, Fuzziness and
  Knowledge-Based Systems}, 10(05):557--570, 2002.

\bibitem{dwork2006calibrating}
Cynthia Dwork, Frank McSherry, Kobbi Nissim, and Adam Smith.
\newblock Calibrating noise to sensitivity in private data analysis.
\newblock In {\em Theory of cryptography conference}, pages 265--284. Springer,
  2006.

\bibitem{sankar2013utility}
Lalitha Sankar, S~Raj Rajagopalan, and H~Vincent Poor.
\newblock Utility-privacy tradeoffs in databases: An information-theoretic
  approach.
\newblock {\em IEEE Transactions on Information Forensics and Security},
  8(6):838--852, 2013.

\bibitem{chim2012vspn}
Tat~Wing Chim, Siu-Ming Yiu, Lucas~CK Hui, and Victor~OK Li.
\newblock Vspn: Vanet-based secure and privacy-preserving navigation.
\newblock {\em IEEE Transactions on Computers}, 63(2):510--524, 2012.

\bibitem{ni2016privacy}
Jianbing Ni, Xiaodong Lin, Kuan Zhang, and Xuemin Shen.
\newblock Privacy-preserving real-time navigation system using vehicular
  crowdsourcing.
\newblock In {\em 2016 IEEE 84th Vehicular Technology Conference (VTC-Fall)},
  pages 1--5. IEEE, 2016.

\bibitem{xu2008real}
Jian-Xin Xu, Sanjib~K Panda, and Tong~Heng Lee.
\newblock {\em Real-time iterative learning control: design and applications}.
\newblock Springer Science \& Business Media, 2008.

\bibitem{ruan2019secure}
M.~Ruan, H.~Gao, and Y.~Wang.
\newblock Secure and privacy-preserving consensus.
\newblock {\em IEEE Trans. Autom. Control}, 64(10):4035--4049, 2019.

\bibitem{zhang2019admm}
Chunlei Zhang, Muaz Ahmad, and Yongqiang Wang.
\newblock Admm based privacy-preserving decentralized optimization.
\newblock {\em IEEE Trans. Inf. Forensic Secur.}, 14(3):565--580, 2019.

\bibitem{Preview1}
E~K.~Bender.
\newblock Optimum linear preview control with application to vehicle
  suspension.
\newblock {\em Journal of Basic Engineering}, 90:213, 01 1968.

\bibitem{Preview2}
N.~Louam, D.~A. Wilson, and R.~S. Sharp.
\newblock Optimal control of a vehicle suspension incorporating the time delay
  between front and rear wheel inputs.
\newblock {\em Vehicle System Dynamics}, 17(6):317--336, 1988.

\bibitem{Preview3}
A.~{Hac} and I.~{Youn}.
\newblock Optimal semi-active suspension with preview based on a quarter car
  model.
\newblock In {\em 1991 American Control Conference}, pages 433--438, June 1991.

\bibitem{roadmarker}
Highway location marker.
\newblock \url{https://en.wikipedia.org/wiki/Highway_location_marker}.
\newblock Accessed: 2020-03-20.

\bibitem{grade}
Narayanan Kidambi, RL~Harne, Yuji Fujii, Gregory~M Pietron, and KW~Wang.
\newblock Methods in vehicle mass and road grade estimation.
\newblock {\em SAE International Journal of Passenger Cars-Mechanical Systems},
  7(2014-01-0111):981--991, 2014.

\bibitem{Profile1}
M.~Doumiati, A.~Victorino, A.~Charara, and D.~Lechner.
\newblock Estimation of road profile for vehicle dynamics motion: Experimental
  validation.
\newblock In {\em Proceedings of the 2011 American Control Conference}, pages
  5237--5242, June 2011.

\bibitem{kalabic2013multi}
Uro{\v{s}} Kalabi{\'c}, Ilya Kolmanovsky, and Julia Buckland.
\newblock Multi-input observer for estimation of compressor flow.
\newblock In {\em ASME 2013 Dynamic Systems and Control Conference}. American
  Society of Mechanical Engineers Digital Collection, 2013.

\bibitem{kolmanovsky2006stochastic}
IV~Kolmanovsky and TL~Maizenberg.
\newblock Stochastic stability, estimation and control in systems driven by
  jump-diffusion disturbances and their automotive applications.
\newblock In {\em Proceedings of the 45th IEEE Conference on Decision and
  Control}, pages 4194--4199. IEEE, 2006.

\bibitem{le2014differentially}
Jerome Le~Ny and George~J Pappas.
\newblock Differentially private filtering.
\newblock {\em IEEE Transactions on Automatic Control}, 59(2):341--354, 2014.

\bibitem{narayanan2008robust}
Arvind Narayanan and Vitaly Shmatikov.
\newblock Robust de-anonymization of large sparse datasets.
\newblock In {\em 2008 IEEE Symposium on Security and Privacy (sp 2008)}, pages
  111--125. IEEE, 2008.

\bibitem{calandrino2011you}
Joseph~A Calandrino, Ann Kilzer, Arvind Narayanan, Edward~W Felten, and Vitaly
  Shmatikov.
\newblock " you might also like:" privacy risks of collaborative filtering.
\newblock In {\em 2011 IEEE symposium on security and privacy}, pages 231--246.
  IEEE, 2011.

\bibitem{wu2009balanced}
Qianhong Wu, Josep Domingo-Ferrer, and Ursula Gonz{\'a}lez-Nicol{\'a}s.
\newblock Balanced trustworthiness, safety, and privacy in vehicle-to-vehicle
  communications.
\newblock {\em IEEE Transactions on Vehicular Technology}, 59(2):559--573,
  2009.

\end{thebibliography}

\begin{IEEEbiographynophoto}{Huan Gao} was born in Shandong, China. He received the B.S. degree in automation and M.Sc. degree in control theory and control engineering from Northwestern Polytechnical University, Shaanxi, China, in 2011 and 2015, respectively. Currently, he is working toward the Ph.D. degree at Clemson University. His research focuses on cooperative control and privacy preservation in distributed systems.
\end{IEEEbiographynophoto}

\begin{IEEEbiographynophoto}{Zhaojian Li} is an Assistant Professor in the Department of Mechanical Engineering at Michigan State University. He obtained M.S. (2013) and Ph.D. (2015) in Aerospace Engineering (flight dynamics and control) at the University of Michigan, Ann Arbor. As an undergraduate, Dr. Li studied at Nanjing University of Aeronautics and Astronautics, Department of Civil Aviation, in China. Dr. Li worked as an algorithm engineer at General Motors from January 2016 to July 2017. His research interests include Learning-based Control, Nonlinear and Complex Systems, and Robotics and Automated Vehicles.
\end{IEEEbiographynophoto}

\begin{IEEEbiographynophoto}{Yongqiang Wang} Yongqiang Wang (Senior Member, IEEE) was born in Shandong, China. He received the B.S. degree in electrical engineering \& automation, the B.S. degree in computer science \& technology from Xian Jiaotong University, Shaanxi, China, in 2004. He received the M.Sc. and the Ph.D. degrees in control science \& engineering from Tsinghua University, Beijing, China, in 2009. From 2007-2008, he was with the University of Duisburg-Essen, Germany, as a visiting student. He was a Project Scientist at the University of California, Santa Barbara before joining Clemson University where he is currently an Associate Professor with the Department of Electrical and Computer Engineering. His research interests are cooperative and networked control, synchronization of wireless sensor networks, systems modeling and analysis of biochemical oscillator networks, and model-based fault diagnosis. 

He currently serves as an associate editor for IEEE Transactions on Control of Network Systems and IEEE Transactions on Signal and Information Processing over Networks. 
\end{IEEEbiographynophoto}

\end{document}